\documentclass[twocolumn]{aastex61}

\shorttitle{The envelope of 47 Tuc}
\shortauthors{Andr\'es E. Piatti}

\begin{document}

\title{Detection of a diffuse extended halo-like structure around 47 Tuc}

\author{Andr\'es E. Piatti}
\affiliation{Consejo Nacional de Investigaciones Cient\'{\i}ficas y T\'ecnicas, Av. Rivadavia 1917, 
C1033AAJ, Buenos Aires, Argentina}
\affiliation{Observatorio Astron\'omico, Universidad Nacional de C\'ordoba, Laprida 854, 5000, 
C\'ordoba, Argentina}
\email{e-mail: andres@oac.unc.edu.ar}

\begin{abstract}

We constructed for the first time a stellar density profile of 47 Tucanae (47 Tuc) out of $\sim$ 5.5 times 
its tidal radius ($r_t$) using high-quality deep $BV$ photometry. After carefully considering the
influence of photometric errors, and Milky Way and Small Magellanic Cloud composite stellar
population contamination, we found that the cluster stellar density profile reaches a
nearly constant value from $\sim$ 1.7$r_t$ outwards, which does not depend on the direction from the
cluster's center considered. These results visibly contrast with recent distinct theoretical predictions on 
the existence of tidal tails or on a density profile that falls as $r^{-4}$ at large distances,
and with observational outcomes of a clumpy structure as well. Our results suggest that the envelope of
47 Tuc is a halo- like nearly constant low density structure.

\end{abstract}


\section{Introduction}

Extended structures around Galactic globular clusters (GGCs) have been observed in a
non-negligible number of objects \citep[e.g. see][]{carballobelloetal2012}. 
\citet{olszewskietal2009} found an unprecedented extra-tidal, azimuthally smooth, 
halo-like diffuse spatial extension of the NGC\,1851, while \citet{correntietal2011} 
discovered an extended stellar halo surrounding the distant NGC\,5694.
M\,2 was also found to be embedded in a diffuse stellar envelope extending to a radial distance 
of at least five time the nominal tidal radius \citep{kuzmaetal2016}.
Compelling evidence of long tidal tails have also been reported in the field of Pal\,5
 \citep{odenetal2003}, Pal\,14 \citep{sollimaetal2011}, Pal\,15 \citep{myeongetal2017},
and NGC\,7492 \citep{naverreteetal2017}, among others. From a theoretical point on view,
N-body simulations  have shown that  the detection of extended envelopes around GGCs
could be due, for instance, to potential escapers \citep{kupperetal2010} or potential observational biases
\citep{bg2017}.

Recent theoretical models argued on very distinct features of the envelope of the
47 Tucanae (47 Tuc). \cite{laneetal2012}  modeled the cluster orbital 
motion to determine the locations and the stellar densities of cluster tidal tails, which predicted 
to be an increase of 3-4\% above the Galactic background. The tails would seem to emerge from the
cluster center towards opposite directions that are connected by a line oriented North-East to
South-West. On the other hand, \cite{penarrubiaetal2017} using statistical arguments and numerical
techniques derived cluster stellar density profiles, assuming that they are embedded in a dark
matter halo. They found that the cluster densities approach asymptotically $\rho \sim r^{-4}$ at
large distances. Models with no dark matter produce much less shallower profiles. 

From an observational point of view, some previous results suggested a clumpy structure around the
cluster \citep{chch2010}. However, they are based on 2MASS photometry that barely reaches the 
cluster's Main Sequence (MS) turnoff region. \citet{leonetal2000} had also pointed out 
the serious challenge that represents the contamination of Small Magellanic Cloud (SMC) stars
that caused they could not trace the cluster radial density profile in direction towards the
 galaxy.

In this Letter we describe how we accomplish constructing a radial stellar density profile of 47 Tuc 
out of $\sim$ 5.5$r_t$ in direction to the SMC and between $\sim$ 1.7 and 3.7$r_t$ for any other
direction from the cluster center. Nevertheless, these outcomes will be greatly benefit, for instance, from the ongoing 
DECam surveys \citep{abbottetal2016}.  In the following we describe the collection and processing
of the data set, and the subsequent analysis performed in order to produce the radial density
profile as a function of the position angle. Finally, we briefly discuss our results.

\section{Data analysis and discussion}

\begin{figure}
\includegraphics[width=\columnwidth]{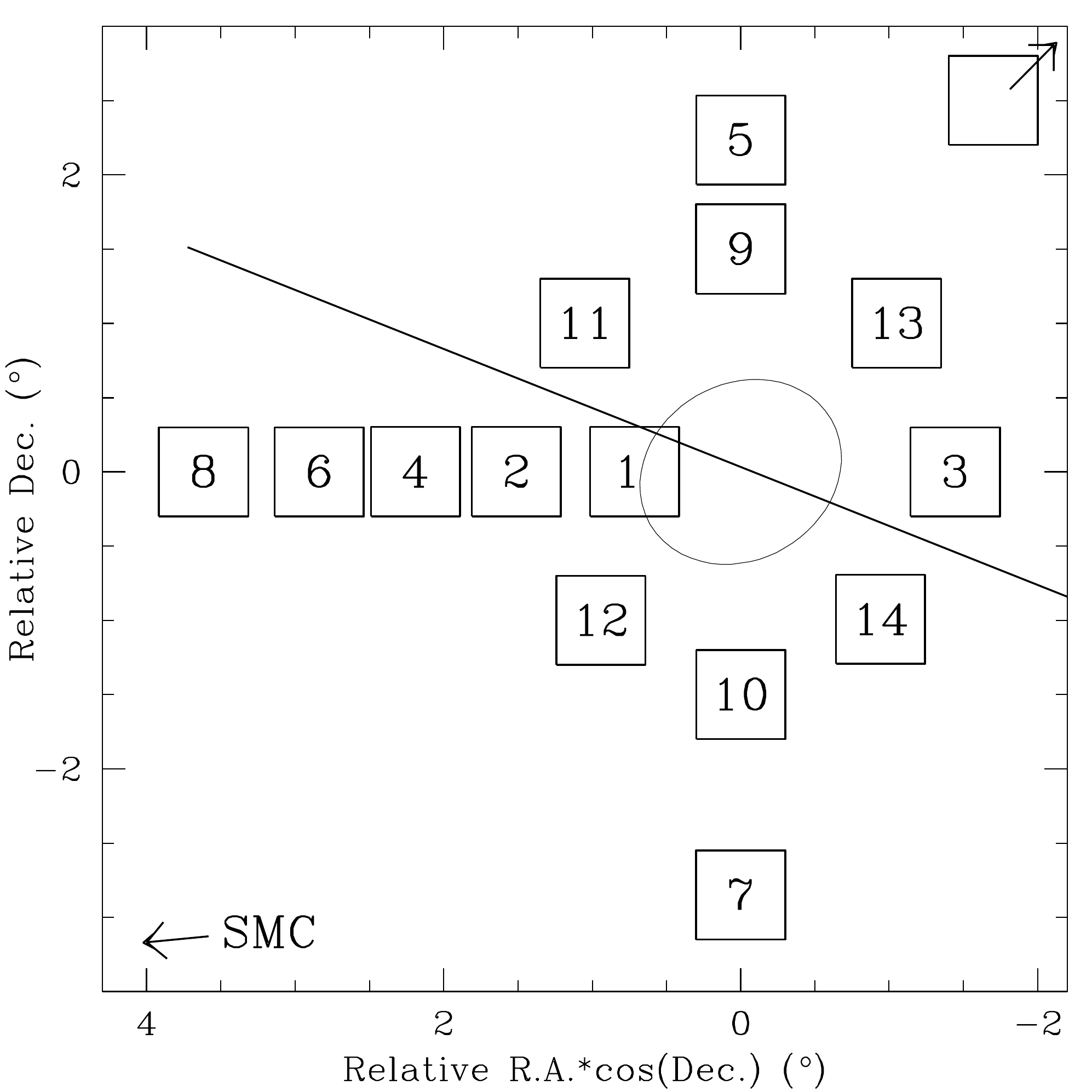}
\caption{Spatial distribution of studied fields ( labeled boxes); the  unlabeled one
refers to two MW fields located $\sim$ 9.3$^o$ from the cluster center. The
ellipse has a semi-major axis equals to the cluster tidal radius (56 pc) and a PA of
120$^o$ \citep{chch2010}. The straight line represents the
position and extension of the tidal tails near the cluster body predicted by \citet{laneetal2012}.
The direction to the SMC is also indicated.}
\label{fig1}
\end{figure}

\begin{figure}
\includegraphics[width=\columnwidth]{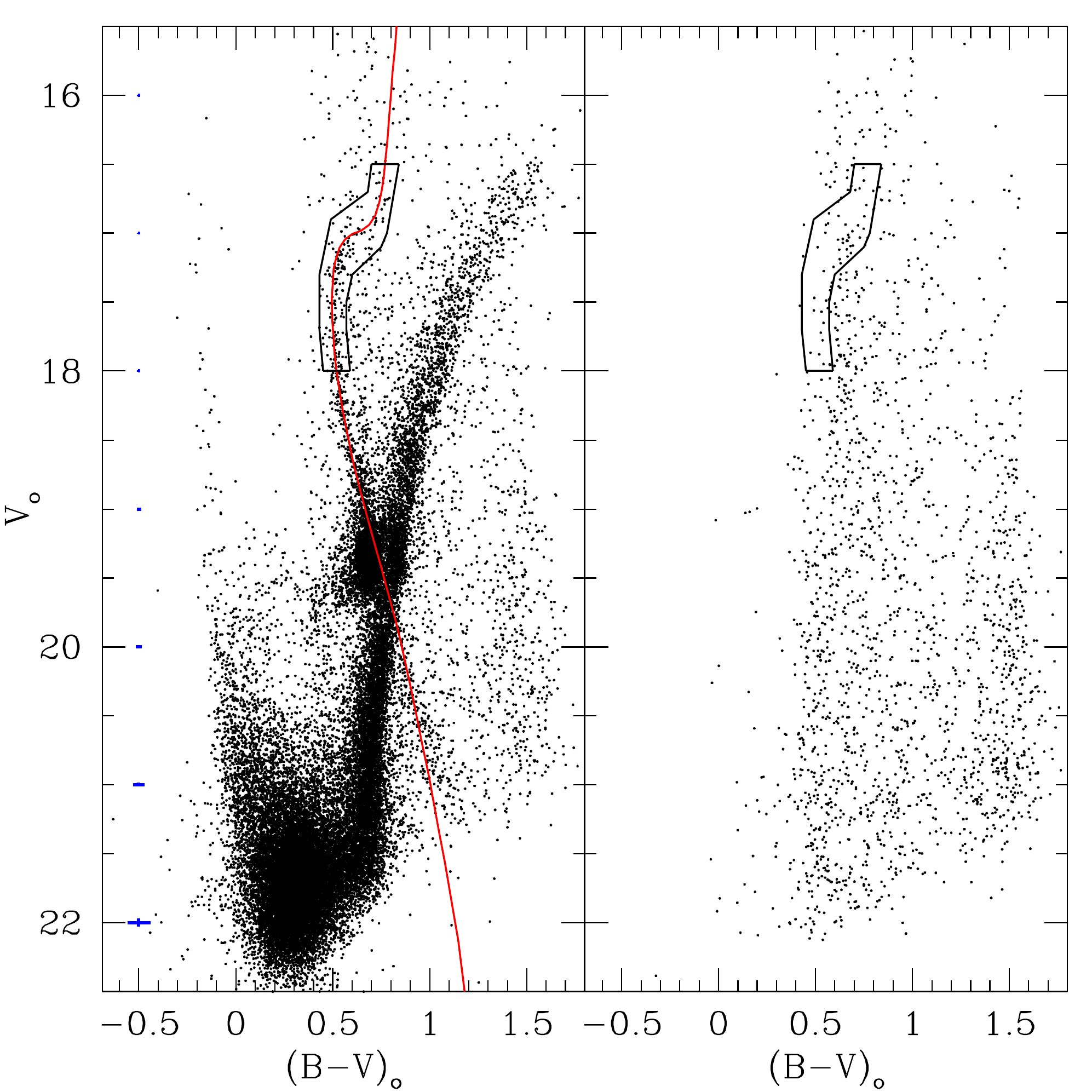}
\caption{Color-magnitude diagram for stars in the 47 Tuc field \# 1 (left panel) and in a MW
field (right panel); errorbars are included at the left margin (blue). The region used
to perform star counts is overplotted. An isochrone of log($t$ yr$^{-1}$) = 10.10, [Fe/H] = -0.7 dex, 
$(m-M)_V$ = 13.37 mag and $E(B-V)$ = 0.033 mag is also superimposed.}
\label{fig2}
\end{figure}

We made use of publicly available 600 s $B$ and 300 s $V$ images obtained at the 4 m Blanco 
telescope (CTIO) with Mosaic\,II (36'$\times$36' camera array) as part of a search for extra tidal 
structure in GGCs (CTIO 2009B-0199, PI= Olszewski). The 14 studied fields
are placed around 47 Tuc (see Fig.~\ref{fig1}, between 1.7 and 5.5 timed its tidal radius 
\citep[= 56 pc,][]{harris1996}, and other two Milky Way (MW) fields are located at $\sim$ 9.3$^o$ to 
the North-West
from the cluster center. This data set, which also includes calibration images (zero, domeflats,
skyflats) and standard field images, was processed as described in \cite[e.g.,][]{p12a,pietal12,p15}.
Mean extinction coefficients of 0.211$\pm$0.024 ($B$) and 0.142$\pm$0.014 ($V$) and color terms
of -0.093$\pm$0.004 ($B$) and 0.038$\pm$0.005 ($V$) were obtained, with rms of 0.030 ($B$) and 
0.027 ($V$). Point-spread-function photometry was performed as extensively described, for instance, in
\citet[][]{petal14,pb16a,pc2017}. Particular success in isolating bona fide stellar objects
was achieved by using roundness values between -0.5 and 0.5 and sharpness values between 0.2 and 1.0.
Errors in $V$ and $B-V$ resulted to be $<$ 0.010 mag for $V$ $<$ 19.0 mag.
Fig.~\ref{fig2} depicts the color-magnitude diagram (CMD)
obtained for stars in the 47 Tuc field \# 1 and for one MW field. The former is dominated by the SMC stellar population, namely the old
MS turnoff, the subgiant and red giant branches, and the red clump superimposed to the
47 Tuc's MS \citep[see e.g.,][]{p12a,p15}. We have superimposed a
theoretical isochrone from \citet{betal12} 
of log($t$ yr$^{-1}$) = 10.10, [Fe/H] = -0.7 dex, $(m-M)_V$ = 13.37 mag and $E(B-V)$ = 0.033 mag
\citep{harris1996}. 

\begin{figure}
\includegraphics[width=\columnwidth]{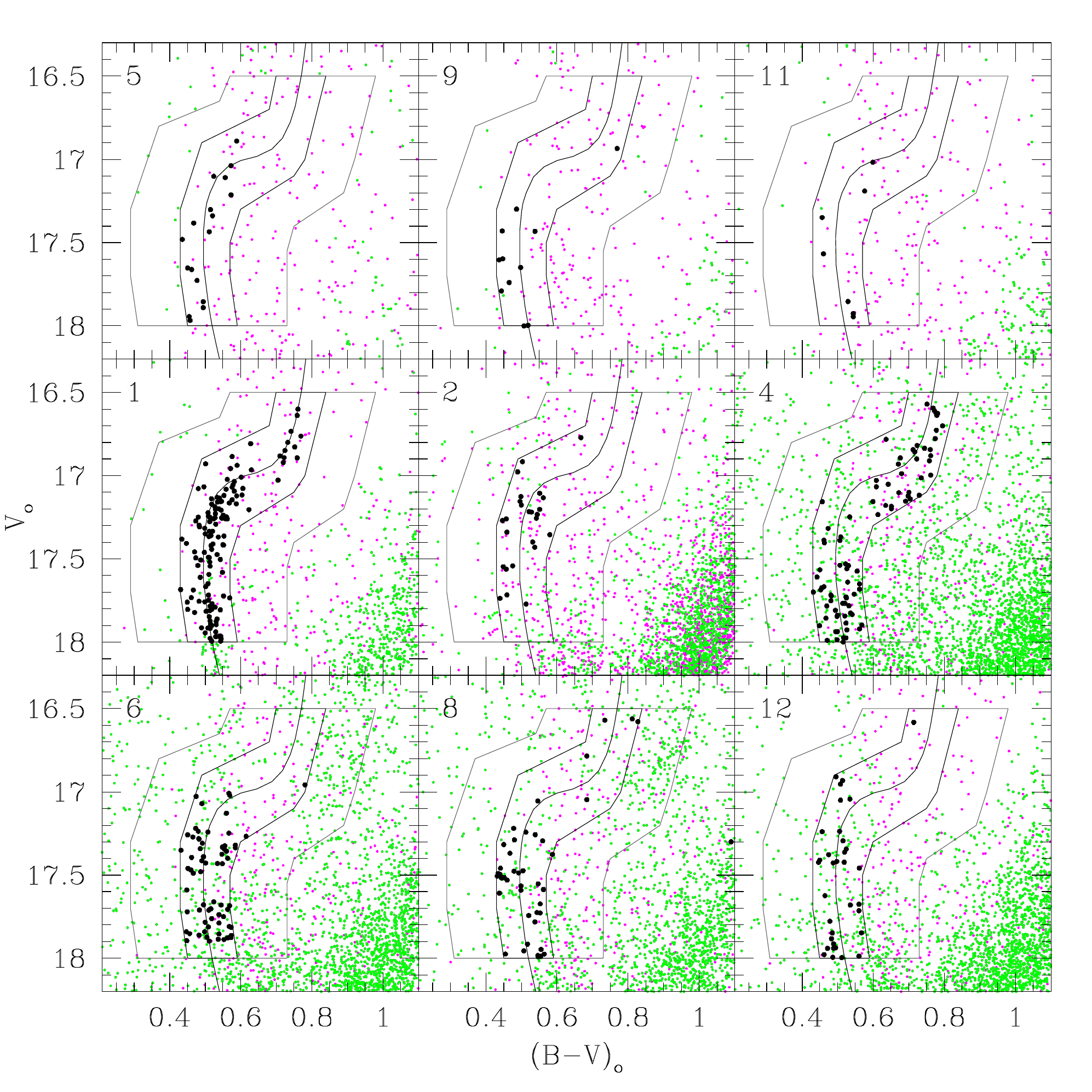}
\includegraphics[width=\columnwidth]{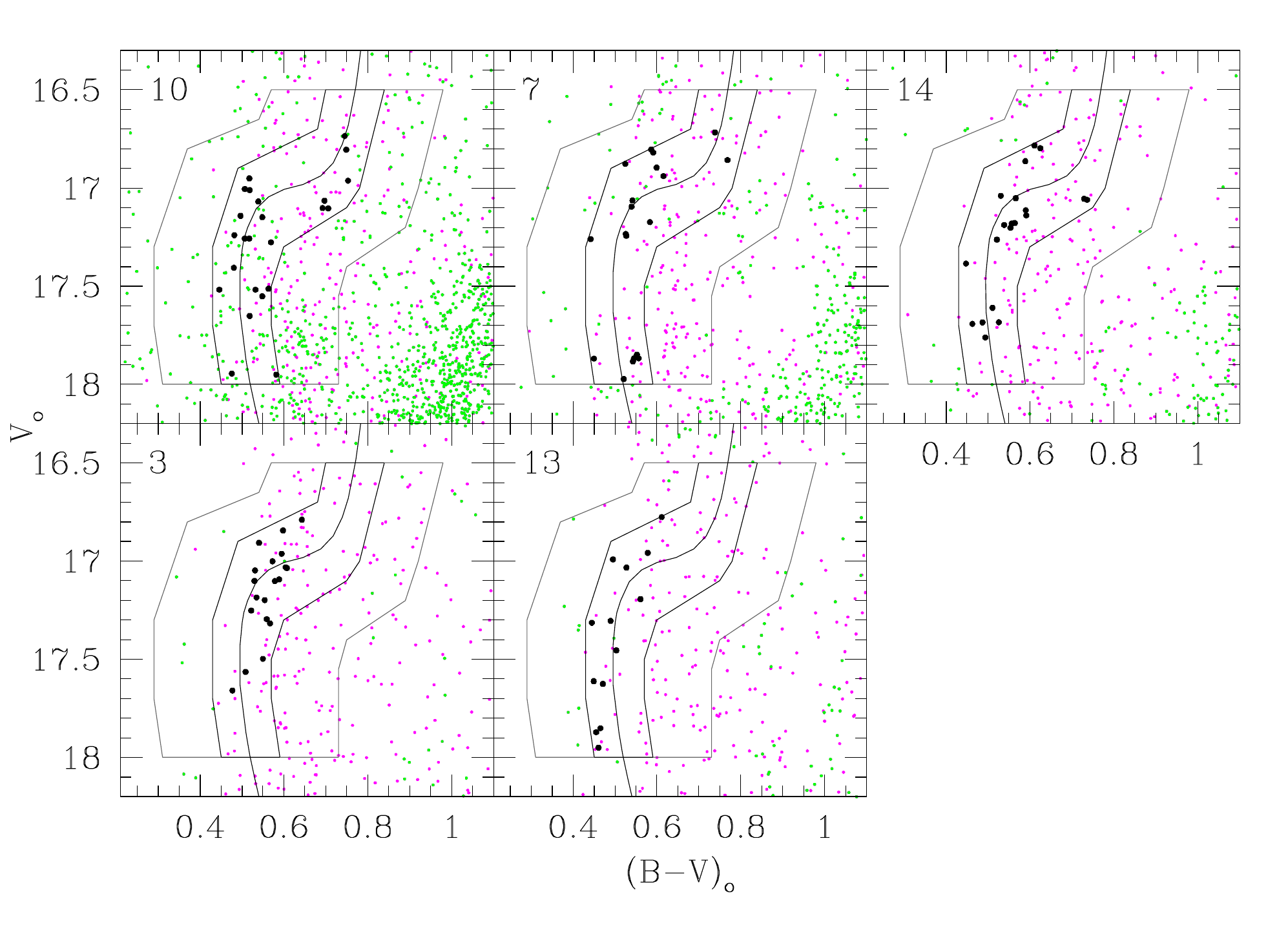}
\caption{Zoomed-in CMDs with statistically subtracted MW (magenta dots), SMC (green dots), and 
47 Tuc (black dots) stars for each studied field ( labeled at the top-left of each panel  and
ordered approximately following increasing position angles (see text for details.)
The regions used to count 47 Tuc (black lines) and SMC ( gray lines) stars are also
 superimposed.}
\label{fig3}
\end{figure}

We dereddened the studied fields using the $E(B-V)$ values as a function of galactic coordinates
obtained by \citet{sf11} from a recalibration of the \citet{schlegeletal1998}'s extinction map. 
The average color excess for the surveyed region is $E(B-V)$ = 0.030$\pm$0.003 mag.
In order to build the cluster density profile, we counted the number of stars distributed inside
 the delineated region drawn in Fig.~\ref{fig2}. The latter comprises the upper cluster
MS and the onset of the subgiant branch, and minimizes the contamination from the SMC.
As for cleaning the observed field CMDs from the MW contamination we applied the procedure outlined by 
\citet{pb12} and successfully used elsewhere \citep[see, e.g.][]{p14,petal15a,p17b}. The MW CMDs
served as the reference field CMD which was subtracted to those observed around 47 Tuc. Statistically
speaking, no residual was left. This is because of the relatively small number of stars in the MW CMD 
and of the uniformity of the MW stellar population throughout the surveyed region. Indeed, we used
two synthetic CMDs generated from the Besan\c{c}on galactic model \citep{retal03}, one centered
on 47 Tuc and the other one at the position of our MW field and, after applying the aforementioned
cleaning precepts, we found none star in the decontaminated CMD. Fig.~\ref{fig3} shows with
magenta dots the observed stars in each 47 Tuc fields that were subtracted using this procedure.
As can be seen, the MW marginally affects the cluster CMD region where we carry out the star
counts.
 
The contamination from the SMC represents a more serious challenge, mainly because its CMD
changes with the position in the sky. Particularly, the region delineated to count cluster stars 
is contamination by supergiants stars, so that the younger (closer to the SMC center) a composite
SMC stellar population, the larger the number of supergiants. In order to cope with this
stellar pollution we used two equal-sized adjacent regions to that traced in Fig.~\ref{fig2}
(see gray contours in Fig~\ref{fig3}). We used these areas to build their
respective luminosity functions, using every star not subtracted previously (those drawn with
green symbols). Then, we adopted the average of both luminosity functions to
subtract the respective number of stars per magnitude interval from the defined
cluster star count region. We used intervals of $\Delta$$V$ = 0.10 mag and subtracted the
appropriate number of stars, randomly. A similar method was employed by \citet{olszewskietal2009}.
The stars that survived this step were drawn with black symbols in Fig.~\ref{fig3}.

We counted the number of measured stars, i.e., stars seen in the observed CMDs without any cleaning procedure,  
distributed along the designed path in the 47 Tuc field CMDs as a function  of  the  distance to the cluster center. 
To do this, we employed the method described by \citep[][and reference therein]{p16b,piattietal2017a},
 based on star counts carried out within statistically  meaningful sized boxes distributed throughout
the whole field, and then computed the number of stars per unit area as a function of the distance $r$ to the
cluster center. This method does not necessarily require a complete circle of radius $r$ within
the observed field to estimate the mean stellar density at that distance. This is an important consideration 
since having a stellar density profile that extends far away from the cluster center allows us to estimate 
it with high precision. We binned  the whole 47 Tuc fields  mosaic (see Fig.~\ref{fig1}) into 
 0.15$^o$$\times$0.15$^o$  boxes.  While performing star counts in the designed 47 Tuc MS strip,
we took into account that a star, owing to its errors, has the chance of falling outside it.
This was done by repeating the star counting with the designed 47 Tuc MS strip shifted in magnitude 
and color by $\pm$ 0.01 mag. We divided Fig.~\ref{fig1} in 8 angular sections of 45$\degr$ wide centered on the cluster,
which resulted suitable for our statistical purposes.

Radial profiles for stars that were kept unsubtracted after cleaning the CMDs from the MW and
SMC field star contamination were also built. In this case, the uncertainties were estimated
taking into account a 20\% fluctuation of the number of stars after cleaning the CMDs from the MW
contamination  ($\sim$ 4 times larger in average than the negligible residuals from MW field star 
variation and cleaning procedure  described above), and twice as large the difference of the number of 
SMC stars subtracted using both 
previously constructed
luminosity functions, in addition to photometric errors. We added in quadrature all the involved 
uncertainties.  Fig.~\ref{fig4} shows the results
with black and magenta circles for observed and cleaned density profiles. We also included the \citet{king62}'s and  \citet{eff87}'s profiles depicted with black and orange lines, respectively, for comparison purposes.

The resultant density profiles along the directions with negligible contamination by SMC stars
(-135$^o$ $\le$ PA $\le$ 45$^o$) and between 70 and 200 pc (1.25 and 3.6$r_t$, respectively)
show  mean stellar excesses of log(stars/deg$^2$) = 1.8$\pm$0.2 . We found slightly
larger values (1.9$\pm$0.3) along the remaining directions (45$^o$ $<$ PA $<$ 225$^o$),
possibly due to residuals of SMC stellar populations. Note that along PA = 90$^o$ our density
profile just starts at the cluster \citet{king62}'s radius and expands until $\sim$ 310 pc (5.5$r_t$).
We recall that these density profiles have been built using mainly upper MS stars, while
fainter MS stars could also unveil these extra tidal structures \citep{carballobelloetal2012},
that have not been used because the SMC overshadows them.
These outcomes suggest that: i) 47 Tuc is not tidally limited to its \citet{king62}'s radius; ii)  
the cluster extends out to at least $\sim$ 5.5$r_t$; iii) from $\sim$ 1.7$r_t$ outwards there is 
a halo- like and nearly constant low density  structure.

We did not find evidence of tidal tails as suggested by \citet{laneetal2012}. According to the authors
they should emerge from the cluster as illustrated by the straight line in Fig.~\ref{fig1} and with
peak stellar densities of $\sim$ 85-120 stars/deg$^2$. Our results show stellar densities 
of the same order as those predicted by the models, though. \citet{chch2010} found a clumpy structure
around the 47 Tuc's center at distances $<$ 250 pc that makes the cluster slightly flattened in
shape (axial ratio of 0.86) with a PA of the major axis of 120$^o$ (see ellipse in Fig.~\ref{fig1}).
They used 2MASS photometry for stars brighter than $K_s$ = 15.6 mag, just barely above
the limiting magnitude of such a database, which in turn nearly coincides with the cluster MS turnoff
magnitude. However, our results suggest that the 47 Tuc's envelope is more likely a diffuse structure,
since the stellar density profiles look similar along any direction from the cluster center. Such
profiles seem to be rather flat, in contrast with the $r^{-4}$ law suggested by \citet{penarrubiaetal2017}
as a prediction of expected stellar envelopes of GCs embedded in dark mini-haloes. \citet{olszewskietal2009}
found a symmetric density profile with a power law of $r^{-1.24}$ profile out of $\sim$ 6$r_t$ in
NGC\,1851, instead, which is more alike to the one derived here for 47 Tuc.
{Although we did not survey uniformly all the sky around 47 Tuc, the present outcome could
suggest that Galactic tidal interactions has been a relatively ineoffficient process for stripping 
stars off the cluster \citep{dinescuetal1997,dinescuetal1999}.

\begin{figure*}
\includegraphics[width=\columnwidth]{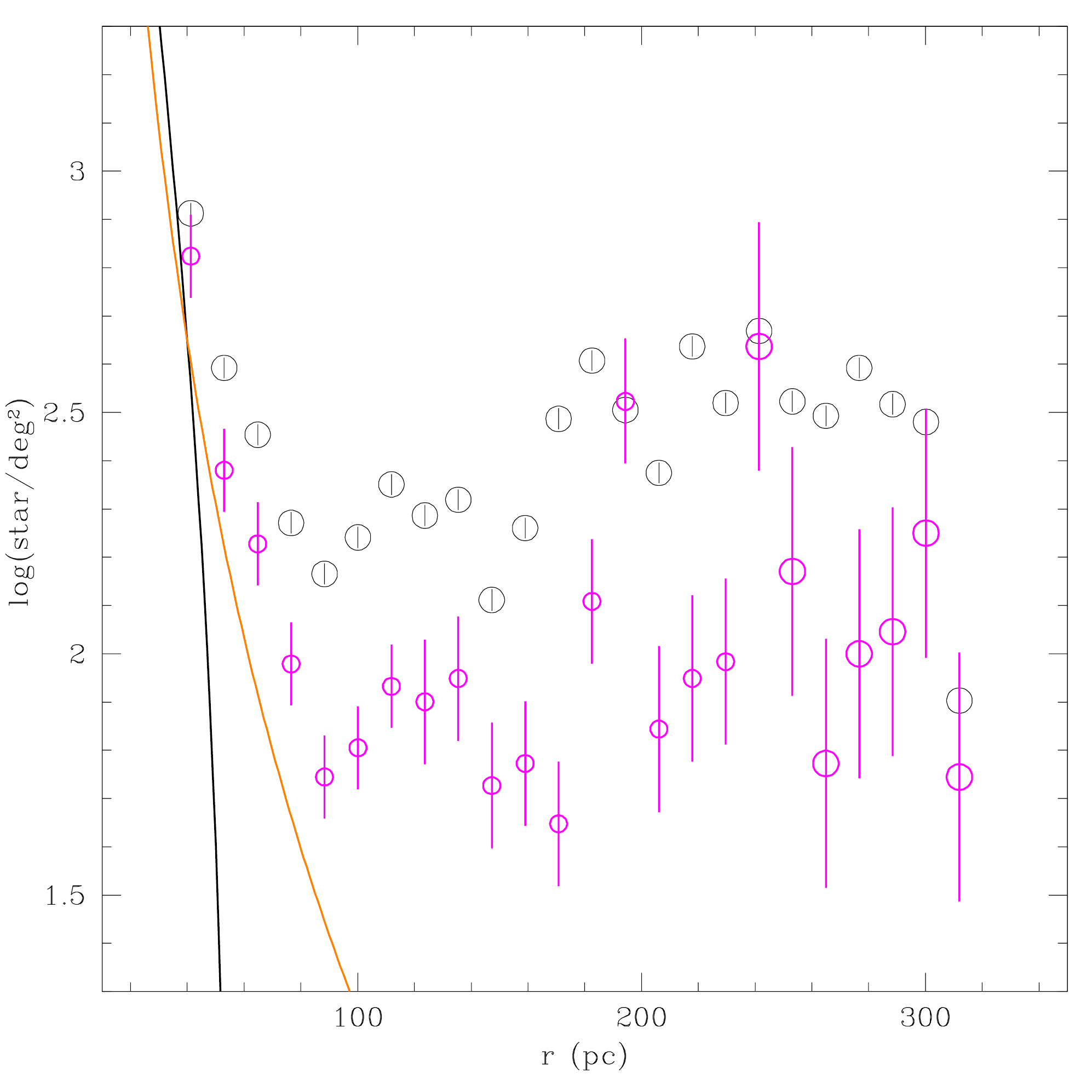}
\includegraphics[width=\columnwidth]{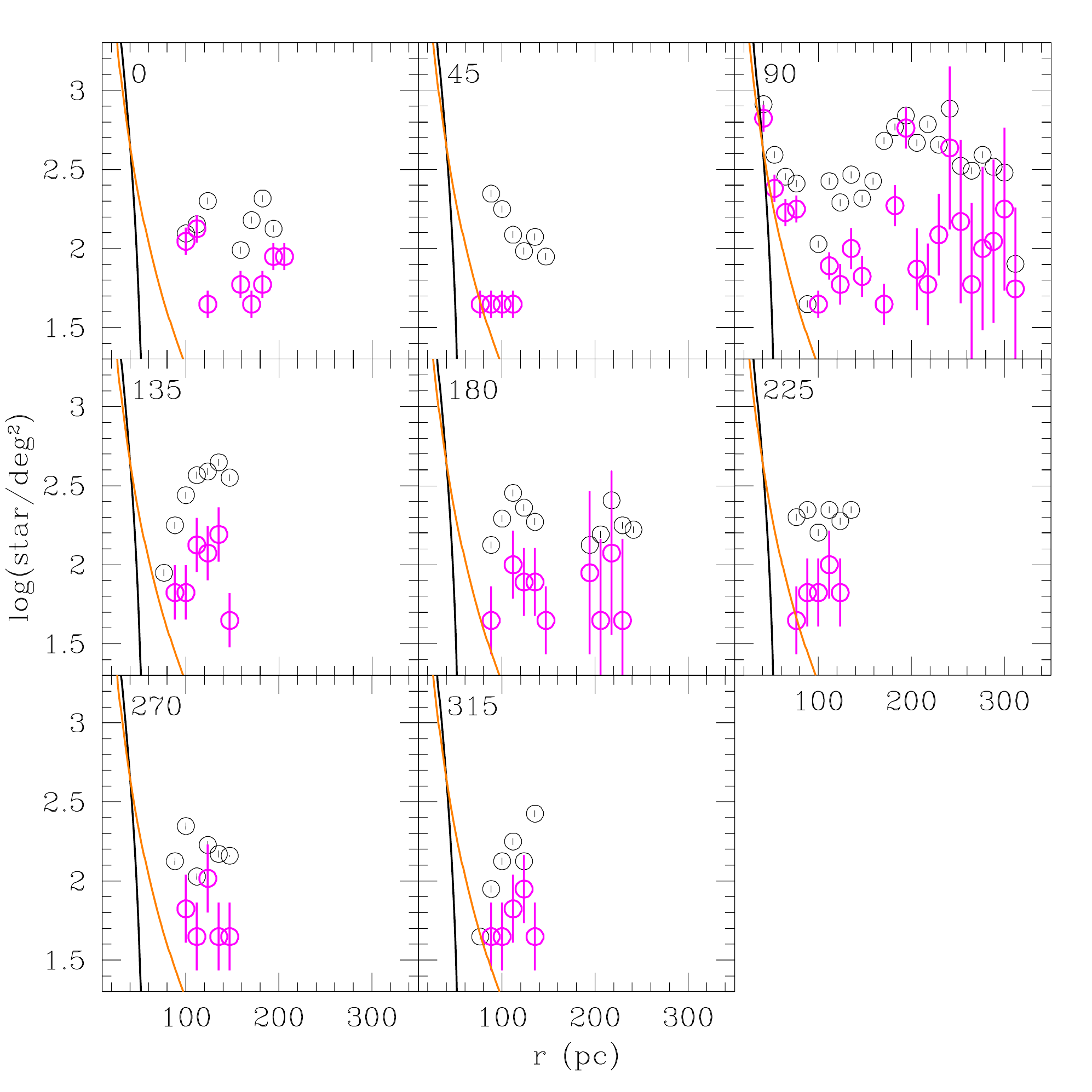}
\caption{Observed (black  circles) and MW-SMC corrected (magenta circles) density
profiles as a function  of the distance to the cluster  center. Central PA
for radial profiles  within PA$\pm$22.5$\degr$ are labeled at the top-left 
margin in degrees  (right).
\citet{king62}'s (black) and  \citet{eff87}'s (orange) models with $r_t$ = 56 pc and $c$ = 2.07 
\citep{harris1996} are also superimposed.  The latter was drawn by adopting a value
of $\gamma$ =3.5 that best resembles the former up to the cluster tidal radius.}
\label{fig4}
\end{figure*}



 We thank the anonymous referee whose thorough comments and suggestions
allowed us to improve the manuscript.

\bibliographystyle{aasjournal}

\begin{thebibliography}{}
\expandafter\ifx\csname natexlab\endcsname\relax\def\natexlab#1{#1}\fi
\providecommand{\url}[1]{\href{#1}{#1}}

\bibitem[{{Abbott} {et~al.}(2016){Abbott}, {Walker}, {Points}, {James},
  {Gregory}, {Tighe}, {David}, {Parkes}, {Cantarutti}, {Warner}, {Estay},
  {Mart{\'{\i}}nez}, {Bonati}, {Bustos}, {Montan{\'e}}, {Mu{\~n}oz}, \&
  {Schurter}}]{abbottetal2016}
{Abbott}, T.~M.~C., {Walker}, A.~R., {Points}, S.~D., {et~al.} 2016, in
  \procspie, Vol. 9906, Ground-based and Airborne Telescopes VI, 99064D

\bibitem[{{Balbinot} \& {Gieles}(2017)}]{bg2017}
{Balbinot}, E., \& {Gieles}, M. 2017, ArXiv e-prints, arXiv:1702.02543

\bibitem[{{Bressan} {et~al.}(2012){Bressan}, {Marigo}, {Girardi}, {Salasnich},
  {Dal Cero}, {Rubele}, \& {Nanni}}]{betal12}
{Bressan}, A., {Marigo}, P., {Girardi}, L., {et~al.} 2012, \mnras, 427, 127

\bibitem[{{Carballo-Bello} {et~al.}(2012){Carballo-Bello}, {Gieles}, {Sollima},
  {Koposov}, {Mart{\'{\i}}nez-Delgado}, \&
  {Pe{\~n}arrubia}}]{carballobelloetal2012}
{Carballo-Bello}, J.~A., {Gieles}, M., {Sollima}, A., {et~al.} 2012, \mnras,
  419, 14

\bibitem[{{Chen} \& {Chen}(2010)}]{chch2010}
{Chen}, C.~W., \& {Chen}, W.~P. 2010, \apj, 721, 1790

\bibitem[{{Correnti} {et~al.}(2011){Correnti}, {Bellazzini}, {Dalessandro},
  {Mucciarelli}, {Monaco}, \& {Catelan}}]{correntietal2011}
{Correnti}, M., {Bellazzini}, M., {Dalessandro}, E., {et~al.} 2011, \mnras,
  417, 2411

\bibitem[{{Dinescu} {et~al.}(1999){Dinescu}, {Girard}, \& {van
  Altena}}]{dinescuetal1999}
{Dinescu}, D.~I., {Girard}, T.~M., \& {van Altena}, W.~F. 1999, \aj, 117, 1792

\bibitem[{{Dinescu} {et~al.}(1997){Dinescu}, {Girard}, {van Altena}, {Mendez},
  \& {Lopez}}]{dinescuetal1997}
{Dinescu}, D.~I., {Girard}, T.~M., {van Altena}, W.~F., {Mendez}, R.~A., \&
  {Lopez}, C.~E. 1997, \aj, 114, 1014

\bibitem[{{Elson} {et~al.}(1987){Elson}, {Fall}, \& {Freeman}}]{eff87}
{Elson}, R.~A.~W., {Fall}, S.~M., \& {Freeman}, K.~C. 1987, \apj, 323, 54

\bibitem[{{Harris}(1996)}]{harris1996}
{Harris}, W.~E. 1996, \aj, 112, 1487

\bibitem[{{King}(1962)}]{king62}
{King}, I. 1962, \aj, 67, 471

\bibitem[{{K{\"u}pper} {et~al.}(2010){K{\"u}pper}, {Kroupa}, {Baumgardt}, \&
  {Heggie}}]{kupperetal2010}
{K{\"u}pper}, A.~H.~W., {Kroupa}, P., {Baumgardt}, H., \& {Heggie}, D.~C. 2010,
  \mnras, 401, 105

\bibitem[{{Kuzma} {et~al.}(2016){Kuzma}, {Da Costa}, {Mackey}, \&
  {Roderick}}]{kuzmaetal2016}
{Kuzma}, P.~B., {Da Costa}, G.~S., {Mackey}, A.~D., \& {Roderick}, T.~A. 2016,
  \mnras, 461, 3639

\bibitem[{{Lane} {et~al.}(2012){Lane}, {K{\"u}pper}, \&
  {Heggie}}]{laneetal2012}
{Lane}, R.~R., {K{\"u}pper}, A.~H.~W., \& {Heggie}, D.~C. 2012, \mnras, 426,
  797

\bibitem[{{Leon} {et~al.}(2000){Leon}, {Meylan}, \& {Combes}}]{leonetal2000}
{Leon}, S., {Meylan}, G., \& {Combes}, F. 2000, \aap, 359, 907

\bibitem[{{Myeong} {et~al.}(2017){Myeong}, {Jerjen}, {Mackey}, \& {Da
  Costa}}]{myeongetal2017}
{Myeong}, G.~C., {Jerjen}, H., {Mackey}, D., \& {Da Costa}, G.~S. 2017, \apjl,
  840, L25

\bibitem[{{Navarrete} {et~al.}(2017){Navarrete}, {Belokurov}, \&
  {Koposov}}]{naverreteetal2017}
{Navarrete}, C., {Belokurov}, V., \& {Koposov}, S.~E. 2017, \apjl, 841, L23

\bibitem[{{Odenkirchen} {et~al.}(2003){Odenkirchen}, {Grebel}, {Dehnen}, {Rix},
  {Yanny}, {Newberg}, {Rockosi}, {Mart{\'{\i}}nez-Delgado}, {Brinkmann}, \&
  {Pier}}]{odenetal2003}
{Odenkirchen}, M., {Grebel}, E.~K., {Dehnen}, W., {et~al.} 2003, \aj, 126, 2385

\bibitem[{{Olszewski} {et~al.}(2009){Olszewski}, {Saha}, {Knezek},
  {Subramaniam}, {de Boer}, \& {Seitzer}}]{olszewskietal2009}
{Olszewski}, E.~W., {Saha}, A., {Knezek}, P., {et~al.} 2009, \aj, 138, 1570

\bibitem[{{Pe{\~n}arrubia} {et~al.}(2017){Pe{\~n}arrubia}, {Varri}, {Breen},
  {Ferguson}, \& {S{\'a}nchez-Janssen}}]{penarrubiaetal2017}
{Pe{\~n}arrubia}, J., {Varri}, A.~L., {Breen}, P.~G., {Ferguson}, A.~M.~N., \&
  {S{\'a}nchez-Janssen}, R. 2017, ArXiv e-prints, arXiv:1706.02710

\bibitem[{{Piatti}(2012)}]{p12a}
{Piatti}, A.~E. 2012, \mnras, 422, 1109

\bibitem[{{Piatti}(2014)}]{p14}
---. 2014, \mnras, 440, 3091

\bibitem[{{Piatti}(2015)}]{p15}
---. 2015, \mnras, 451, 3219

\bibitem[{{Piatti}(2016)}]{p16b}
---. 2016, \mnras, 463, 3476

\bibitem[{{Piatti}(2017)}]{p17b}
---. 2017, \mnras, 465, 2748

\bibitem[{{Piatti} \& {Bastian}(2016)}]{pb16a}
{Piatti}, A.~E., \& {Bastian}, N. 2016, \aap, 590, A50

\bibitem[{{Piatti} \& {Bica}(2012)}]{pb12}
{Piatti}, A.~E., \& {Bica}, E. 2012, \mnras, 425, 3085

\bibitem[{{Piatti} \& {Cole}(2017)}]{pc2017}
{Piatti}, A.~E., \& {Cole}, A. 2017, ArXiv e-prints, arXiv:1705.08186

\bibitem[{{Piatti} {et~al.}(2015){Piatti}, {de Grijs}, {Rubele}, {Cioni},
  {Ripepi}, \& {Kerber}}]{petal15a}
{Piatti}, A.~E., {de Grijs}, R., {Rubele}, S., {et~al.} 2015, \mnras, 450, 552

\bibitem[{{Piatti} {et~al.}(2017){Piatti}, {Dias}, \&
  {Sampedro}}]{piattietal2017a}
{Piatti}, A.~E., {Dias}, W.~S., \& {Sampedro}, L.~M. 2017, \mnras, 466, 392

\bibitem[{{Piatti} {et~al.}(2012){Piatti}, {Geisler}, \& {Mateluna}}]{pietal12}
{Piatti}, A.~E., {Geisler}, D., \& {Mateluna}, R. 2012, \aj, 144, 100

\bibitem[{{Piatti} {et~al.}(2014){Piatti}, {Keller}, {Mackey}, \& {Da
  Costa}}]{petal14}
{Piatti}, A.~E., {Keller}, S.~C., {Mackey}, A.~D., \& {Da Costa}, G.~S. 2014,
  \mnras, 444, 1425

\bibitem[{{Robin} {et~al.}(2003){Robin}, {Reyl{\'e}}, {Derri{\`e}re}, \&
  {Picaud}}]{retal03}
{Robin}, A.~C., {Reyl{\'e}}, C., {Derri{\`e}re}, S., \& {Picaud}, S. 2003,
  \aap, 409, 523

\bibitem[{{Schlafly} \& {Finkbeiner}(2011)}]{sf11}
{Schlafly}, E.~F., \& {Finkbeiner}, D.~P. 2011, \apj, 737, 103

\bibitem[{{Schlegel} {et~al.}(1998){Schlegel}, {Finkbeiner}, \&
  {Davis}}]{schlegeletal1998}
{Schlegel}, D.~J., {Finkbeiner}, D.~P., \& {Davis}, M. 1998, \apj, 500, 525

\bibitem[{{Sollima} {et~al.}(2011){Sollima}, {Mart{\'{\i}}nez-Delgado},
  {Valls-Gabaud}, \& {Pe{\~n}arrubia}}]{sollimaetal2011}
{Sollima}, A., {Mart{\'{\i}}nez-Delgado}, D., {Valls-Gabaud}, D., \&
  {Pe{\~n}arrubia}, J. 2011, \apj, 726, 47

\end{thebibliography}

\end{document}